\documentstyle[12pt]{article}
\begin{document}
\baselineskip8mm
\begin{titlepage}
\title{\vspace{-3.5cm}
{\bf Casimir effect in problems with spherical symmetry: new
perspectives}}
\author{Giampiero Esposito$^{1}$, Alexander Yu. Kamenshchik$^{2}$ \\
and Klaus Kirsten$^{3}$}
\maketitle
\hspace{-6mm}$^{1}${\em Istituto Nazionale di Fisica Nucleare,
Sezione di Napoli, Mostra d'Oltremare Padiglione 20, 
80125 Napoli, Italy}\\
$^{2}${\em L. D. Landau Institute for Theoretical Physics of
Russian Academy of Sciences, Kosygina Str. 2, 
Moscow 117334, Russia}\\
$^{3}${\em U\-ni\-ver\-si\-t\"{a}t 
Le\-ip\-zig, In\-sti\-tut f\"{u}r The\-o\-re\-tis\-che
Phy\-sik, Au\-gu\-stus\-pla\-tz 10, 
0\-41\-09 Le\-ip\-zig, Ger\-ma\-ny}

{\bf Abstract.}
Since the Maxwell theory of electromagnetic phenomena
is a gauge theory, it is quite important to evaluate the 
zero-point energy of the quantized electromagnetic field by a 
careful assignment of boundary conditions on the potential and
on the ghost fields. Recent work by the authors has shown that,
for a perfectly conducting spherical shell, it is precisely the
contribution of longitudinal and normal modes of the potential 
which enables one to reproduce the result first due to Boyer. This
is obtained provided that one works with the Lorenz gauge-averaging
functional, and with the help of the Feynman choice for a 
dimensionless gauge parameter. For arbitrary values of the gauge
parameter, however, covariant and non-covariant gauges lead to an
entangled system of three eigenvalue equations. Such a problem is
crucial both for the foundations and for the applications of
quantum field theory.
\vfill
\end{titlepage}

\def\cstok#1{\leavevmode\thinspace\hbox{\vrule\vtop{\vbox{\hrule\kern1pt
\hbox{\vphantom{\tt/}\thinspace{\tt#1}\thinspace}}
\kern1pt\hrule}\vrule}\thinspace}

Since Casimir produced his remarkable calculation of zero-point
energy of the electromagnetic field for the case of two perfectly
conducting plates of area $A$ and separation $d$:
${\bigtriangleup E}=-{\pi^{2}{\hbar}c \over 720}{A\over d^{3}}$,
there has been an increasing series of efforts in the literature to
evaluate suitable differences in zero-point energies for various
sorts of geometries and in a variety of media. Since this is a 
research field where the theoretical predictions have been tested
against observations \cite{Lamor97,Mohid98}, 
it is clear why the investigation of Casimir energies
has attracted efforts over half a century, with even better
perspectives for the years to come. Theoretical physicists, however,
who are also interested in the general structures, may approach the
Casimir energy calculation with the aim of re-deriving a non-trivial
prediction in a way which is viewed as more fundamental from the
point of view of general principles. In particular, we refer here
to the path-integral quantization of gauge theories. As is well known,
on using the Faddeev-Popov formalism, one performs Gaussian averages
over gauge functionals $\chi^{\mu}$, by adding to the original
Lagrangian a gauge-averaging term $\chi^{\mu}\beta_{\mu \nu}\chi^{\nu}$,
where $\beta_{\mu \nu}$ is any constant invertible matrix. This turns 
the original operator on field perturbations into a new operator 
which has the advantage of being non-degenerate. The result of a
one-loop calculation is expected to be $\chi$- and $\beta$-independent,
although no rigorous proof exists on manifolds with boundary. In
particular, for Maxwell theory, $\beta_{\mu \nu}$ reduces to a
$1 \times 1$ matrix, i.e. a real-valued parameter, and $\chi^{\mu}$
reduces to the familiar covariant or non-covariant gauges for
Maxwell theory, e.g. Lorenz, Coulomb, axial.

Our paper describes recent work by the authors 
\cite{Espo98} on the application of 
the Lorenz and axial gauges to the evaluation of the zero-point 
energy of a perfectly conducting spherical shell. The gauge-invariant
boundary conditions applied by Boyer in his seminal 
paper \cite{Boyer68} require 
that tangential components of the electric field should vanish
on a two-sphere of radius $R$. More precisely, on denoting by
$r,\theta, \varphi$ the spherical coordinates, which
are appropriate for the analysis of a conducting spherical
shell, one writes
\begin{equation}
[E_{\theta}]_{\partial M}=0 ,
\label{1}
\end{equation}
\begin{equation}
[E_{\varphi}]_{\partial M}=0 .
\label{2}
\end{equation}
The modes of the
electromagnetic field are then split into transverse electric
(TE) or magnetic multipole:
$
{\vec r} {\cdot} {\vec E}=E_{r}=0 ,
$
and transverse magnetic (TM) or electric multipole:
$
{\vec r} {\cdot} {\vec B}=B_{r}=0 .
$
On using the standard notation for spherical Bessel
functions, one finds that, in the TE case, Eqs. (\ref{1})
and (\ref{2}) lead to
\begin{equation}
\Bigr[A_{l} j_{l}(kr)+B_{l} n_{l}(kr)\Bigr]_{\partial M}=0,
\label{3}
\end{equation}
while in the TM case the vanishing of $E_{\theta}$ and 
$E_{\varphi}$ at the boundary implies that
\begin{equation}
\left[{d\over dr}(r(C_{l} j_{l}(kr)+D_{l} n_{l}(kr)))
\right]_{\partial M}=0 .
\label{4}
\end{equation}
Of course, if the background includes the point $r=0$, the
coefficients $B_{l}$ and $D_{l}$ should be set to zero for
all values of $l$ to obtain a regular solution. Such a 
singularity is instead avoided if one studies the annular
region in between two concentric spheres.

On the other hand, if one follows a path-integral
approach to the quantization of Maxwell theory, it is 
well known that the second-order operator acting on
$A^{\mu}$ perturbations when the Lorenz gauge-averaging 
functional is chosen turns out to be, in a flat background,
\begin{equation}
P_{\mu \nu}=-g_{\mu \nu} \cstok{\ } 
+\left(1-{1\over \alpha} \right) \nabla_{\mu} \nabla_{\nu} .
\label{5}
\end{equation}
With a standard notation, $g$ is the background metric, 
$\alpha$ is a dimensionless parameter,
$\cstok{\ }$ is the D'Alembert operator
$$
\cstok{\ } \equiv -{\partial^{2}\over \partial t^{2}}
+ \bigtriangleup ,
$$
where $\bigtriangleup$ is the Laplace operator 
(in our problem, $\bigtriangleup$ is considered on a disk). 
This split of the $\cstok{\ }$ operator leads eventually to the
eigenvalue equations for the Laplace operator acting on the
temporal, normal and tangential components of the potential 
(see below).
At this stage, the potential, with its gauge transformations
\begin{equation}
{ }^{\varepsilon}A_{\mu} \equiv A_{\mu}+
\nabla_{\mu} \varepsilon ,
\label{6}
\end{equation}
is viewed as a more fundamental object. Some care,
however, is then necessary to ensure gauge invariance of the
whole set of boundary conditions. For example, if one imposes
the boundary conditions
\begin{equation}
[A_{t}]_{\partial M}=0 ,
\label{7}
\end{equation}
\begin{equation}
[A_{\theta}]_{\partial M}=0 ,
\label{8}
\end{equation}
\begin{equation}
[A_{\varphi}]_{\partial M}=0 ,
\label{9}
\end{equation}
this is enough to ensure that the boundary conditions (\ref{1})
and (\ref{2}) hold. However, within
this framework, one has to impose yet another  
condition. By virtue of Eq. (\ref{6}), the desired boundary
condition involves the gauge function:
\begin{equation}
[\varepsilon ]_{\partial M}=0 .
\label{10}
\end{equation}
Equation (\ref{10}) ensures that the boundary conditions
(\ref{7})--(\ref{9}) are preserved under the gauge transformations
(\ref{6}). What happens is that $\varepsilon$ is expanded in
harmonics on the two-sphere according to the relation
\begin{equation}
\varepsilon(t,r,\theta,\varphi)=\sum_{l=0}^{\infty}
\sum_{m=-l}^{l} \varepsilon_{l}(r) Y_{lm}(\theta,\varphi)
e^{i \omega t} .
\label{11}
\end{equation}
After a gauge transformation, one finds from (\ref{6})
\begin{equation}
[{ }^{\varepsilon}A_{t}]_{\partial M}-[A_{t}]_{\partial M}
=[\nabla_{t}\varepsilon]_{\partial M} ,
\label{12}
\end{equation}
\begin{equation}
[{ }^{\varepsilon}A_{\theta}]_{\partial M}
-[A_{\theta}]_{\partial M}
=[\nabla_{\theta}\varepsilon]_{\partial M},
\label{13}
\end{equation}
\begin{equation}
[{ }^{\varepsilon}A_{\varphi}]_{\partial M}
-[A_{\varphi}]_{\partial M}
=[\nabla_{\varphi}\varepsilon]_{\partial M},
\label{14}
\end{equation}
and by virtue of (\ref{11}) one has
\begin{equation}
[\nabla_{t}\varepsilon]_{\partial M}
=i \omega [\varepsilon]_{\partial M},
\label{15}
\end{equation}
\begin{equation}
[\nabla_{\theta}\varepsilon]_{\partial M}
=\left[\sum_{l=0}^{\infty} \sum_{m=-l}^{l} \varepsilon_{l}(r)
Y_{lm,\theta}(\theta,\varphi)e^{i\omega t}\right]_{\partial M},
\label{16}
\end{equation}
\begin{equation}
[\nabla_{\varphi}\varepsilon]_{\partial M}
=\left[\sum_{l=0}^{\infty} \sum_{m=-l}^{l} \varepsilon_{l}(r)
Y_{lm,\varphi}(\theta,\varphi)e^{i\omega t}\right]_{\partial M}.
\label{17}
\end{equation}
Thus, if $\varepsilon_{l}(r)$ is set to zero at the boundary 
$\forall l$, the right-hand sides of (\ref{15})--(\ref{17}) vanish at
$\partial M$. But
$$
[\varepsilon_{l}(r)]_{\partial M}=0 \; \forall l
$$
is precisely the condition which ensures the vanishing of
$\varepsilon$ at $\partial M$, and the proof of our statement
is completed.

At this stage, the only boundary condition on
$A_{r}$ whose preservation under the transformation (\ref{6})
is again guaranteed by Eq. (\ref{10}) is the vanishing of the
gauge-averaging functional $\Phi(A)$ at the boundary:
\begin{equation}
[\Phi(A)]_{\partial M}=0 .
\label{18}
\end{equation}
On choosing the Lorenz term $\Phi_{L}(A) \equiv
\nabla^{b}A_{b}$, Eq. (\ref{18}) leads to 
\begin{equation}
{\left[{\partial A_{r} \over \partial r}
+{2\over r}A_{r} 
\right]}_{\partial M}=0 .
\label{19}
\end{equation}
It is only upon considering the joint effect of Eqs. 
(\ref{7})--(\ref{10}), (\ref{18}) and (\ref{19})  
that the whole set of boundary conditions
becomes gauge-invariant. This scheme is also BRST-invariant.
In the following calculations, it will be
enough to consider a {\it real-valued gauge function} obeying Eq.
(\ref{10}) at the boundary, and then multiply the resulting
contribution to the zero-point energy by $-2$, bearing in
mind the fermionic nature of ghost fields for the electromagnetic
field.

In our problem, the electromagnetic potential has a temporal
component $A_{t}$, a normal component $A_{r}$, and tangential
components $A_{k}$ (hereafter, 
$k$ refers to $\theta$ and $\varphi$). They can all
be expanded in harmonics on the two-sphere, according to the
standard formulae
\begin{equation}
A_{t}(t,r,\theta,\varphi)=\sum_{l=0}^{\infty}\sum_{m=-l}^{l}
a_{l}(r)Y_{lm}(\theta,\varphi)e^{i\omega t},
\label{20}
\end{equation}
\begin{equation}
A_{r}(t,r,\theta,\varphi)=\sum_{l=0}^{\infty}\sum_{m=-l}^{l}
b_{l}(r)Y_{lm}(\theta,\varphi)e^{i\omega t},
\label{21}
\end{equation}
\begin{equation}
A_{k}(t,r,\theta,\varphi)=\sum_{l=1}^{\infty}\sum_{m=-l}^{l}
\Bigr[c_{l}(r)\partial_{k}Y_{lm}(\theta,\varphi)
+T_{l}(r)\varepsilon_{kp}\partial^{p}Y_{lm}(\theta,\varphi)
\Bigr]e^{i\omega t}.
\label{22}
\end{equation}
This means that we are performing a Fourier analysis 
of the components of the electromagnetic potential.
The two terms in square brackets of Eq. (\ref{22}) refer to
longitudinal and transverse modes, respectively.
On setting $\alpha=1$ in Eq. (\ref{5}), the evaluation of
basis functions for electromagnetic perturbations can be
performed after studying the action of the operator
$-g_{\mu\nu}\cstok{\ }$ on the components (\ref{20})--(\ref{22}).
For this purpose, we perform the analytic continuation
$\omega \rightarrow iM$, which makes it possible to express
the basis functions in terms of modified Bessel functions
(of course, one could work equally well with $\omega$,
which leads instead to ordinary Bessel functions). For the
temporal component, one deals with 
the Laplacian acting on a scalar field on the 
three-dimensional disk:
\begin{equation}
({ }^{(3)}\bigtriangleup A)_{t}={\partial^{2}A_{t} \over
\partial r^{2}}+{2\over r}{\partial A_{t}\over \partial r}
+{1\over r^{2}} ({ }^{(2)}\bigtriangleup A)_{t} ,
\label{23}
\end{equation}
which leads to the eigenvalue equation 
\begin{equation}
\left[{d^{2}\over dr^{2}}+{2\over r}{d\over dr}
-{l(l+1)\over r^{2}}\right]a_{l}=M^{2}a_{l} .
\label{24}
\end{equation}
Hereafter, we omit for simplicity any subscript for $M^{2}$,
and the notation for the modes will make it sufficiently
clear which spectrum is studied.
The solution of Eq. (\ref{24}) which is regular at $r=0$
is thus found to be
\begin{equation}
a_{l}(r)={1\over \sqrt{r}}I_{l+1/2}(Mr),
\label{25}
\end{equation}
up to an unessential multiplicative constant.

The action of ${ }^{(3)} \bigtriangleup$ on 
the component $A_{r}$ normal to the two-sphere is
\begin{equation}
({ }^{(3)}\bigtriangleup A)_{r}=
{\partial^{2}A_{r}\over \partial r^{2}}+{2\over r}
{\partial A_{r}\over \partial r}
+{1\over r^{2}}({ }^{(2)}\bigtriangleup A)_{r}
-{2\over r^{2}}A_{r}
-{2\over r^{3}}A_{p}^{\; \; \mid p} ,
\label{26}
\end{equation}
where the stroke $\mid$ denotes two-dimensional covariant
differentiation on a two-sphere of unit radius. Last, the
Laplacian on tangential components takes the form
\begin{equation}
({ }^{(3)}\bigtriangleup A)_{k}=
{\partial^{2}A_{k}\over \partial r^{2}}
+{1\over r^{2}}({ }^{(2)}\bigtriangleup A)_{k}
-{1\over r^{2}}A_{k}+{2\over r}\partial_{k}A_{r} .
\label{27}
\end{equation}
By virtue of the expansions (\ref{21}) and (\ref{22}), jointly with
the standard properties of spherical harmonics, Eqs.
(\ref{26}) and (\ref{27}) lead to the eigenvalue equation
\begin{equation}
\left[{d^{2}\over dr^{2}}-{l(l+1)\over r^{2}}
\right]T_{l}=M^{2}T_{l}
\label{28}
\end{equation}
for transverse modes, jointly with entangled eigenvalue
equations for normal and longitudinal modes
(here $l \geq 1$):
\begin{equation}
\left[{d^{2}\over dr^{2}}+{2\over r}{d\over dr}
-{(l(l+1)+2)\over r^{2}}\right]b_{l}
+{2l(l+1)\over r^{3}}c_{l}=M^{2} b_{l},
\label{29}
\end{equation}
\begin{equation}
\left[{d^{2}\over dr^{2}}-{l(l+1)\over r^{2}}
\right]c_{l}+{2\over r}b_{l}=M^{2} c_{l}.
\label{30}
\end{equation}
The mode $b_{0}(r)$ is instead decoupled, and is proportional
to ${I_{3/2}(Mr)\over \sqrt{r}}$ in the interior problem.
It is indeed well known that gauge modes of the Maxwell
field obey a coupled set of eigenvalue equations. In
arbitrary gauges, one cannot decouple these modes. This
can be proved by trying to put in diagonal form the
$2 \times 2$ operator matrix acting on the modes $b_{l}$
and $c_{l}$. In our problem, however,
with our choice of gauge-averaging functional and gauge
parameter, gauge modes can be disentangled, and a simpler
method to achieve this exists. For this purpose, we point
out that, since the background is flat, if gauge modes can
be decoupled, they can only reduce to linear combinations 
of Bessel functions, i.e.,
\begin{equation}
b_{l}(r)={B_{\nu}(Mr)\over \sqrt{r}},
\label{31}
\end{equation}
and
\begin{equation}
c_{l}(r)=C(\nu)B_{\nu}(Mr)\sqrt{r} .
\label{32}
\end{equation}
With our notation, $C(\nu)$ is some constant depending on
$\nu$, which is obtained in turn from $l$. To find $\nu$ 
and $C(\nu)$, we insert the ansatz (\ref{31}) and (\ref{32}) into the
system of equations (\ref{29}) and (\ref{30}), and we require that 
the resulting equations should be of Bessel type for 
$B_{\nu}(Mr)$, i.e.,
\begin{equation}
\left[{d^{2}\over dr^{2}}+{1\over r}{d\over dr}-M^{2}
\left(1+{\nu^{2}\over M^{2}r^{2}}\right)\right]
B_{\nu}(Mr)=0.
\label{33}
\end{equation}
This leads to two algebraic equations for $\nu^{2}$.
By comparison, one thus finds an algebraic equation of second
degree for C:
\begin{equation}
l(l+1)C^{2}-C-1=0,
\label{34}
\end{equation}
whose roots are
$C_{+}={1\over l}$,
and
$C_{-}=-{1\over (l+1)}$.
The corresponding values of $\nu$ are
$\nu_{+}=l-{1\over 2}$, and
$\nu_{-}=l+{3\over 2}$.
Hence one finds the basis functions for normal and 
longitudinal perturbations in the interior problem in the form
\begin{equation}
b_{l}(r)=\alpha_{1,l}{I_{l+3/2}(Mr)\over \sqrt{r}}
+\alpha_{2,l}{I_{l-1/2}(Mr)\over \sqrt{r}},
\label{35}
\end{equation}
\begin{equation}
c_{l}(r)=-{\alpha_{1,l}\over (l+1)}
I_{l+3/2}(Mr)\sqrt{r}+{\alpha_{2,l}\over l}
I_{l-1/2}(Mr)\sqrt{r},
\label{36}
\end{equation}
whereas, from Eq. (\ref{28}), transverse modes read
\begin{equation}
T_{l}(r)=I_{l+1/2}(Mr)\sqrt{r}.
\label{37}
\end{equation}
Last, but not least, ghost modes obey an eigenvalue equation 
analogous to (\ref{24}), i.e. 
\begin{equation}
\left[{d^{2}\over dr^{2}}+{2\over r}{d\over dr}
-{l(l+1)\over r^{2}}\right]\varepsilon_{l}=M^{2}\varepsilon_{l},
\label{38}
\end{equation}
and hence they read
\begin{equation}
\varepsilon_{l}(r)={1\over \sqrt{r}}I_{l+1/2}(Mr).
\label{39}
\end{equation}
In the exterior problem, i.e., for $r$ greater than the two-sphere
radius $R$, one has simply to replace the modified Bessel functions
of first kind in Eqs. (\ref{25}), (\ref{35})--(\ref{37})
and (\ref{39}) by modified Bessel
functions of second kind, to ensure regularity at infinity.

In our problem, ghost modes are of course decoupled from the 
modes for the electromagnetic potential which occur in the 
expansions (\ref{20})--(\ref{22}). Nevertheless, this does not mean that
they do not play a role in the Casimir energy calculation. By
contrast, we already know (see comments after (\ref{10}))
that boundary conditions on the ghost are strictly necessary to
ensure gauge invariance of the boundary conditions on the potential. 
It is then clear that such boundary conditions, combined with the
differential equation (\ref{38}), lead to a ghost spectrum whose
contribution to the Casimir energy can only be obtained after a
detailed calculation (e.g. Green-function approach, or
$\zeta$-function regularization). It should not
be surprising that ghost terms are important, since one
has already included effects of other degrees of freedom which
should be compensated for \cite{Espo98}. 
This issue is further clarified
by the analysis of the original Casimir problem: two perfectly
conducting parallel plates. In a covariant formalism, one has to
consider the energy-momentum tensor for ghosts, which is found to
give a non-vanishing contribution to the average energy density
in vacuum. After taking into account 
boundary conditions for the potential entirely analogous to our
Eqs. (\ref{7})--(\ref{9}) and (\ref{18}), 
one then finds a renormalized value
of the zero-point energy in complete agreement with the result first
found by Casimir.

By virtue of Eq. (\ref{7}), the modes $a_{l}(r)$ obey homogeneous
Dirichlet conditions:
\begin{equation}
[a_{l}(r)]_{\partial M}=0, \forall l \geq 0.
\label{40}
\end{equation}
Moreover, Eq. (\ref{19}) implies that the modes $b_{l}$ obey the
boundary conditions
\begin{equation}
\left[{\partial \over \partial r} r^{2}b_{l}(r)
\right]_{\partial M}=0, \forall l \geq 0.
\label{41}
\end{equation}
Last, the modes $c_{l}$ and $T_{l}$, being the tangential
modes, obey Dirichlet boundary conditions (cf. Eqs. (\ref{8})
and (\ref{9}))
\begin{equation}
[c_{l}(r)]_{\partial M}=[T_{l}(r)]_{\partial M}=0,
\forall l \geq 1.
\label{42}
\end{equation}
On taking into account how the modes are expressed in
terms of Bessel functions (see Eqs. 
(\ref{25}), (\ref{35})--(\ref{37})),
one thus finds five sets of eigenvalue conditions
for the interior and exterior problems, respectively:
\vskip 0.3cm
\noindent
(i) Temporal modes: 
\begin{equation}
I_{l+1/2}(MR)=0, \forall l \geq 0,
\label{43}
\end{equation}
\begin{equation}
K_{l+1/2}(MR)=0, \forall l \geq 0.
\label{44}
\end{equation}
\vskip 0.3cm
\noindent
(ii) Decoupled normal mode:
\begin{equation}
\left[{\partial \over \partial r}r^{3/2}I_{3/2}(Mr)
\right]_{r=R}=0,
\label{45}
\end{equation}
\begin{equation}
\left[{\partial \over \partial r}r^{3/2}K_{3/2}(Mr)
\right]_{r=R}=0.
\label{46}
\end{equation}
\vskip 0.3cm
\noindent
(iii) Coupled longitudinal and normal modes (here
$\nu \equiv l + 3/2$):
\begin{eqnarray}
\; & \; &(\nu-1/2)I_{\nu}'(MR)I_{\nu -2}(MR)
+3(\nu-1){I_{\nu}(MR)\over MR}I_{\nu-2}(MR) \nonumber \\
&+& (\nu-3/2)I_{\nu-2}'(MR)I_{\nu}(MR)=0,
\label{47}
\end{eqnarray}
\begin{eqnarray}
\; & \; & (\nu-1/2)K_{\nu}'(MR)K_{\nu -2}(MR)
+3(\nu-1){K_{\nu}(MR)\over MR}K_{\nu-2}(MR) \nonumber \\
&+& (\nu-3/2)K_{\nu-2}'(MR)K_{\nu}(MR)=0.
\label{48}
\end{eqnarray}
\vskip 0.3cm
\noindent
(iv) Transverse modes:
\begin{equation}
I_{l+1/2}(MR)=0, \forall l \geq 1,
\label{49}
\end{equation}
\begin{equation}
K_{l+1/2}(MR)=0, \forall l \geq 1.
\label{50}
\end{equation}
\vskip 0.3cm
\noindent
(v) Ghost modes (multiplying their $\zeta$-function by -2):
\begin{equation}
I_{l+1/2}(MR)=0, \forall l \geq 0,
\label{51}
\end{equation}
\begin{equation}
K_{l+1/2}(MR)=0, \forall l \geq 0.
\label{52}
\end{equation}

The eigenvalue conditions (\ref{45}) and 
(\ref{46}) can be re-expressed in the form
\begin{equation}
I_{1/2}(MR)=0,
\label{53}
\end{equation}
\begin{equation}
K_{1/2}(MR)=0.
\label{54}
\end{equation}
It is thus clear that, by construction, the contribution of Eqs.
(\ref{53}) and (\ref{54}) to the Casimir energy of a conducting spherical
shell cancels exactly the joint effect of Eqs. (\ref{43}), (\ref{44}),
(\ref{49}), (\ref{50}), (\ref{51}) and (\ref{52}), 
bearing in mind the fermionic
nature of ghost fields. In general, each set of boundary conditions
involving a set of positive eigenvalues $\left \{ \lambda_{k}
\right \}$ contributes to the Casimir energy in a way which is
clarified by the $\zeta$-function method, because the regularized
ground-state energy is defined by the equation (for ${\rm {Re}}(s)
> s_{0}=2$)
\begin{equation}
E_{0}(s) \equiv -{1\over 2} \sum_{k} (\lambda_{k})^{{1\over 2}-s}
\; \mu^{2s}=-{1\over 2} \zeta \left(s-{1\over 2} \right)\mu^{2s},
\label{55}
\end{equation}
which is later analytically continued to the value $s=0$ in the
complex-$s$ plane. Here $\mu$ is the usual mass parameter and $\zeta$
is the $\zeta$-function of the positive-definite elliptic operator 
$\cal B$ with discrete spectrum $\left \{ \lambda_{k} \right \}$:
\begin{equation}
\zeta_{{\cal B}}(s) \equiv {\rm {Tr}}_{L^{2}}({\cal B}^{-s})
=\sum_{k} (\lambda_{k})^{-s}.
\label{56}
\end{equation}
In other words, the regularized ground-state energy is equal to
$-{1\over 2}\zeta_{{\cal B}}(-1/2)$, at least for the cases where
$\zeta_{{\cal B}}(s)$ has no pole at $s=-1/2$, as is the case
for the problem considered here. In general, however, this will
not be true but instead one has $\mbox{Res}\,\, \zeta_{{\cal B}}
(-1/2) \sim b_{2}$, and as a result an ambiguity for the
ground-state energy proportional to the heat-kernel coefficient
$b_{2}$ of the operator ${\cal B}$ remains. In these cases a
renormalization procedure has to be performed to eliminate 
this ambiguity. 

Although the eigenvalues are
known only implicitly, the form of the function occurring in the
mode-by-mode expression of the boundary conditions leads eventually
to $E_{0}(0)$. The non-trivial part of the analysis is 
represented by Eqs. (\ref{47}) and (\ref{48}).
These are obtained by imposing the Robin boundary
conditions for normal modes, and Dirichlet conditions for 
longitudinal modes. To find non-trivial solutions of the resulting
linear systems of two equations in the unknowns $\alpha_{1,l}$
and $\alpha_{2,l}$, the determinants of the matrices of coefficients
should vanish. This leads to Eqs. (\ref{47}) 
and (\ref{48}). At this stage,
it is more convenient to re-express such equations in terms of 
Bessel functions of order $l+1/2$. On using the standard recurrence
relations among Bessel functions and their first derivatives, one
thus finds the following equivalent forms of eigenvalue conditions:
\begin{equation}
I_{l+1/2}(MR)\left[I_{l+1/2}'(MR)+{1\over 2MR}
I_{l+1/2}(MR)\right]=0, \forall l \geq 1,
\label{57}
\end{equation}
\begin{equation}
K_{l+1/2}(MR)\left[K_{l+1/2}'(MR)+{1\over 2MR}
K_{l+1/2}(MR)\right]=0, \forall l \geq 1.
\label{58}
\end{equation}
Thus, the contribution of the coupled normal and longitudinal
modes splits into the sum of contributions of two scalar fields
obeying Dirichlet and Robin boundary conditions, respectively,
with the $l=0$ mode omitted. This corresponds exactly to the
contributions of TE and TM modes (Eqs. (\ref{3}) and (\ref{4})), 
and gives the same contribution as the one found by Boyer
\cite{Boyer68}.

Following our initial remarks, it is now 
quite important to understand the
key features of the Casimir-energy calculations in other gauges.
For this purpose, we consider a gauge-averaging functional of the
axial type, i.e.
\begin{equation}
\Phi(A) \equiv n^{\mu}A_{\mu},
\label{59}
\end{equation}
where $n^{\mu}$ is the unit normal vector field  
$n^{\mu}=(0,1,0,0)$. The resulting gauge-field operator is
found to be, in our flat background,
\begin{equation}
P^{\mu \nu}=-g^{\mu \nu}\cstok{\ } + \nabla^{\mu} \nabla^{\nu}
+{1\over \alpha}n^{\mu}n^{\nu}.
\label{60}
\end{equation}
Note that, unlike the case of Lorenz gauge, the $\alpha$
parameter is dimensionful and has dimension $[\rm{length}]^{2}$.
Now we impose again the boundary condition according to which the
gauge-averaging functional should vanish at $\partial M$:
\begin{equation}
[\Phi(A)]_{\partial M}=[n^{\mu}A_{\mu}]_{\partial M}
=[A_{r}]_{\partial M}=0,
\label{61}
\end{equation}
which implies that all $b_{l}$ modes vanish at the boundary
(cf. Eq. (\ref{41})). 

A further consequence of the axial gauge may be derived by acting
on the field equations 
\begin{equation}
P^{\mu \nu}A_{\nu}=0
\label{62}
\end{equation}
with the operations of covariant differentiation and contraction
with the unit normal, i.e. 
\begin{equation}
\nabla_{\mu}P^{\mu \nu}A_{\nu}=0,
\label{63}
\end{equation}
\begin{equation}
n_{\mu}P^{\mu \nu}A_{\nu}=0.
\label{64}
\end{equation}
Equation (\ref{63}) leads, in flat space, to the differential equation
\begin{equation}
{\partial A_{r}\over \partial r}+{2\over r}A_{r}=0.
\label{65}
\end{equation}
This first-order equation leads to $b_{l}$ modes having the form
\begin{equation}
b_{l}={b_{0,l}\over r^{2}}.
\label{66}
\end{equation}
Thus, by virtue of the boundary condition (\ref{61}), the modes
$b_{l}$ vanish everywhere, and hence $A_{r}$ vanishes
identically if the axial gauge-averaging functional is chosen with
such boundary conditions.

Moreover, Eq. (\ref{64}) leads to the equation
\begin{equation}
-i\omega {\partial A_{t}\over \partial r}
+{1\over r^{2}}{\partial \over \partial r}
A_{k}^{\; \mid k}=0,
\label{67}
\end{equation}
which implies, upon making the analytic continuation
$\omega \rightarrow iM$, 
\begin{equation}
M{da_{l}\over dr}-{l(l+1)\over r^{2}}{dc_{l}\over dr}=0.
\label{68}
\end{equation}
At this stage, the transverse modes $T_{l}$ obey again the
eigenvalue equation (\ref{28}), whereas the remaining set of modes
obey differential equations which, from Eq. (\ref{62}), are found
to be
\begin{equation}
\left[{d^{2}\over dr^{2}}+{2\over r}{d\over dr}
-{l(l+1)\over r^{2}}\right]a_{l}-M{l(l+1)\over r^{2}}c_{l}=0,
\label{69}
\end{equation}
\begin{equation}
{d^{2}c_{l}\over dr^{2}}-Ma_{l}-M^{2}c_{l}=0,
\label{70}
\end{equation}
\begin{equation}
a_{0}=0.
\label{71}
\end{equation}
In particular, Eq. (\ref{71}) is obtained from Eq. (\ref{70})
(when $l=0$), which is
a reduced form of the Eq. $P_{k\nu}A^{\nu}=0$ upon bearing in
mind that $b_{l}$ modes vanish everywhere. 

Last, but not least, one should consider the ghost operator, which,
in the axial gauge, is found to be 
\begin{equation}
Q=-{\partial \over \partial r}.
\label{72}
\end{equation}
This leads to ghost modes having the form
\begin{equation}
\varepsilon_{l}=\varepsilon_{0,l} \; e^{-Mr}.
\label{73}
\end{equation}
On the other hand, following the method described after 
Eq. (\ref{10}),
one can prove that, also in the axial gauge, the ghost field
should vanish at the boundary, to ensure gauge invariance of the
whole set of boundary conditions on the potential. It is then
clear, from Eq. (\ref{73}), that ghost modes vanish everywhere in
the axial gauge.

On studying the system (\ref{68})--(\ref{70}) one has first to prove 
that these three equations are compatible. This is indeed the
case, because differentiation with respect to $r$ of Eq. (\ref{68})
leads to a second-order equation which, upon expressing
${dc_{l}\over dr}$ from Eq. (\ref{68}) and ${d^{2}c_{l}\over dr^{2}}$
from Eq. (\ref{70}), is found to coincide with Eq. (\ref{69}). 
Thus, we have a system of two second-order differential
equations for two functions $a_{l}$ and $c_{l}$. However, these
functions are not independent, in that they are connected by
Eq. (\ref{68}). Hence for every value of $l$ one has one degree of
freedom instead of two. Finally, we have for every $l$ two degrees
of freedom, one resulting from Eqs. (\ref{68})--(\ref{70}), and another, 
i.e. the transverse mode $T_{l}$. 
Thus, an estimate of the number of degrees of freedom which
contribute to the Casimir energy coincides with that in other gauges.
Moreover, the parameter $\alpha$ does not affect the Casimir 
energy, since $\alpha$ does not occur in any of the eigenvalue
equations.
 
Unfortunately, we cannot obtain the exact form of the solutions
of Eqs. (\ref{69}) and (\ref{70}) in terms of special functions
(e.g. Bessel or hypergeometric). This crucial point can be made
precise by remarking that, if it were possible to disentangle
the system (\ref{69}) and (\ref{70}), one could find some functions 
$\alpha_{l}, \beta_{l}, V_{l}, W_{l}$ such that the
$2 \times 2$ matrix 
$$
\pmatrix{1 & V_{l} \cr W_{l} & 1 \cr}
\pmatrix{{\hat A}_{l} & {\hat B}_{l} \cr
{\hat C}_{l} & {\hat D}_{l} \cr}
\pmatrix{1 & \alpha_{l} \cr \beta_{l} & 1 \cr}
$$
has no non-vanishing off-diagonal elements, where the operators
${\hat A}_{l}, {\hat B}_{l}, {\hat C}_{l}, {\hat D}_{l}$ are the
ones occurring in Eqs. (\ref{69}) and (\ref{70}). For
example, the first off-diagonal element of such matrix is the 
operator 
${\hat A}_{l} \alpha_{l}+{\hat B}_{l}+V_{l}({\hat C}_{l}\alpha_{l}
+{\hat D}_{l})$.
On setting to zero the coefficients of ${d^{2}\over dr^{2}}$ and
${d\over dr}$ one finds that $\alpha_{l}=-V_{l}
={\alpha_{0,l}\over r}$, where $\alpha_{0,l}$ is a constant. But
it is then impossible to set to zero the ``potential" term of this
operator, i.e. its purely multiplicative part.

Nevertheless, there is some evidence that the axial gauge may be
consistently used to evaluate the Casimir energy. For this
purpose we find it helpful to consider a simpler problem, i.e.
the Casimir energy in the axial gauge for the
case of flat boundary. In this case the basis functions are plane
waves
\begin{equation}
A_{\mu}=A_{0,\mu}e^{i(k_{x}x+k_{y}y+k_{z}z-\omega t)},
\label{74}
\end{equation}  
where the admissible values of $k_{x}, k_{y}, k_{z}$ are
determined by the boundary conditions, which are here taken
to be analogous to the case of a curved boundary. Let us choose
the conducting boundaries parallel to the $x$- and $y$-axes,
while the vector $n^{\mu}$ is directed along the $z$-axis.
Then the condition of compatibility of field equations in the
axial gauge is reduced to 
\begin{equation}
D \equiv {\rm {det}} 
\pmatrix{-k^{2} & \omega k_{x} & \omega k_{y} & \omega k_{z} \cr
\omega k_{x} & k_{y}^{2}+k_{z}^{2}-\omega^{2} &
-k_{x} k_{y} & -k_{x} k_{z} \cr
\omega k_{y} & -k_{x} k_{y} & k_{x}^{2}+k_{z}^{2}-\omega^{2}
& -k_{y} k_{z} \cr
\omega k_{z} & -k_{x}k_{z} & -k_{y} k_{z} 
& k_{x}^{2}+k_{y}^{2}-\omega^{2}+{1\over \alpha} \cr}=0.
\label{75}
\end{equation}
Direct calculation shows that this determinant is equal to
\begin{equation}
D=-{1\over \alpha}k_{z}^{2}\left(\omega^{2}-k^{2}\right)^{2}.
\label{76}
\end{equation}
Hence we have reproduced the correct dispersion relation
between energy $\omega$ and wave number $k$ (to every admissible
value of $k$ there correspond two contributions to the Casimir
energy of the form $\omega = \mid {\vec k} \mid$). Of course,
no non-vanishing ghost modes exist, once that the axial type
gauge-averaging functional is set to zero at the boundary
(cf. (\ref{61}), (\ref{72}) and (\ref{73})). Interestingly, on imposing
the boundary conditions, one can set to zero the $A_{z}$
component of the electromagnetic potential as was done with
$A_{r}$ in the spherical case. In this case the compatibility
condition of field equations is reduced to the vanishing of the
determinant of a $3 \times 3$ matrix, obtained by omitting the
fourth row and the fourth column in Eq. (\ref{75}). One can easily
see that the determinant of this $3 \times 3$ matrix coincides
with that of the $4 \times 4$ matrix up to a multiplicative
factor ${1\over \alpha}$, which does not affect the dispersion
relation. 

To sum up, a complete correspondence can be established between
the key features of the axial gauge in the cases of flat and
curved boundary: the $\alpha$ parameter does not affect the Casimir
energy, the component of the potential orthogonal to the boundary
vanishes, ghost modes vanish, and only two independent degrees
of freedom contribute. On going from the flat to the curved case,
however, the analysis of the dispersion relation is replaced by
the problem of finding explicit solutions 
of Eqs. (\ref{68})--(\ref{70}),
with the corresponding eigenvalue conditions. This last technical
problem goes beyond the present capabilities of the authors, but
the exact results and complete correspondences established so far
seem to add evidence in favour of a complete solution being in
sight in the near future.

We have studied an approach to the evaluation of the
zero-point energy of a conducting spherical shell which relies
on a careful investigation of the potential and of ghost fields,
with the corresponding set of boundary conditions on $A_{\mu}$
perturbations and ghost modes. When Boyer first developed his
calculation, the formalism of ghost fields for the quantization
of gauge fields and gravitation had just been developed, and
hence it is quite natural that, in the first series of papers on
Casimir energies, the ghost contribution was not considered, 
since the emphasis was always put on TE and TM modes for the
electromagnetic field. On the other hand, the Casimir energy is a
peculiar property of the quantum theory, and an approach via
path-integral quantization regards the potential and the ghost as
more fundamental. This is indeed necessary to take into account
that Maxwell theory is a gauge theory. 
Some of these issues had been studied in the literature
\cite{Ambjo83,Bordag85},
including calculations with ghosts in covariant gauges,
but, to our knowledge, an explicit mode-by-mode analysis of the
$A_{\mu}$ and ghost contributions in problems with spherical
symmetry was still lacking in the 
literature. The contributions of our investigation are as follows
\cite{Espo98}.

First, the basis functions for the electromagnetic potential have
been found explicitly when the Lorenz gauge-averaging functional
is used. The temporal, normal, longitudinal and transverse modes
have been shown to be linear combinations of spherical Bessel
functions. Second, it has been proved that transverse modes of the
potential are, by themselves, unable to
reproduce the correct value for the Casimir energy of a conducting
spherical shell. Third, it is exactly the effect of coupled 
longitudinal and normal modes of $A_{\mu}$ which is responsible for
the value of $\bigtriangleup E=0.09 {\hbar}c/2R$ found
by Boyer. This adds evidence in favour 
of physical degrees of freedom for gauge theories being a concept
crucially depending on the particular problem under consideration
and on the boundary conditions. Fourth, ghost modes play a
non-trivial role as well, in that they cancel the contribution 
resulting from transverse, decoupled and temporal 
modes of the potential. Fifth, the axial gauge-averaging functional
has been used to study the Casimir energy for Boyer's problem.
Unlike the case of the Lorenz gauge, ghost modes and normal
modes are found to vanish, and one easily proves that the result
is independent of the $\alpha$ parameter. A complete comparison
with the case of flat boundary has also been performed, getting 
insight into the problems of independent degrees of freedom
and non-vanishing contributions to the Casimir energy. 

Indeed, recent investigations of Euclidean Maxwell theory in 
quantum cosmological backgrounds had already shown that
longitudinal, normal and ghost modes are all essential to
obtain the value of the conformal anomaly and of the one-loop
effective action \cite{Espo97}. Further evidence of the
non-trivial role played by ghost modes in curved backgrounds
had been obtained, much earlier, in Ref. 
\cite{Shore79}. Hence, we find it
non-trivial that the ghost formalism 
gives results in complete agreement with Boyer's investigation.
Note also that, in the case of
perfectly conducting parallel plates, the ghost contribution
cancels the one due to tangential components of the potential
(see Sec. 4.5 of Ref. \cite{Espo97}). 
This differs from the cancellations
found in our paper in the presence of spherical symmetry.

The main open problem is now the explicit proof that the Casimir
energy is independent of the choice of
gauge-averaging functional. This task is as difficult as
crucial to obtain a thorough understanding of the quantized
electromagnetic field in the presence of bounding surfaces.
In general, one has then to study entangled eigenvalue
equations for temporal, normal and longitudinal modes.
The general solution is not (obviously)
expressed in terms of well known special functions. 
A satisfactory understanding of this 
problem is still lacking, while its solution would be of great
relevance both for the foundations and for the applications of
quantum field theory. In covariant gauges, coupled eigenvalue
equations with arbitrary gauge parameter also lead to severe
technical problems, which are described in Ref. \cite{Espo98}.
A further set of non-trivial applications lies
in the consideration of more involved geometries, where spherical
symmetry no longer holds, in the investigation of media which
are not perfect conductors, and in the analysis of 
radiative corrections \cite{Grib94,Bordag98}. 
Thus, there exists increasing evidence
that the study of Casimir energies will continue to play
an important role in quantum field theory in the years to come,
and that a formalism relying on potentials and ghost fields is,
indeed, crucial on the way
towards a better understanding of quantized fields.

\end{document}